\documentclass[12pt,a4paper,british]{iopart}
\usepackage{mathptmx}

\usepackage[T1]{fontenc}
\usepackage[latin9]{inputenc}
\usepackage{amstext}
\usepackage{graphicx}
\usepackage{esint}

\makeatletter

\newcommand{\noun}[1]{\textsc{#1}}
\providecommand{\tabularnewline}{\\}

\usepackage{iopams}
\usepackage{setstack}

\@ifundefined{definecolor}

\usepackage{dcolumn}
\usepackage{epsfig}
\usepackage{epstopdf}
\usepackage{babel}

\makeatother

\usepackage{babel}
\begin{document}

\title{Calibration and High Fidelity Measurement of a Quantum Photonic Chip}

\author{H W Li$^{1}$, J Wabnig$^{1}$, D Bitauld$^{1}$, P Shadbolt$^{2}$, A Politi$^{2,3}$, A Laing$^{2}$, J L O'Brien$^{2}$, and A O Niskanen$^{1}$}

\address{$^{1}$ Nokia Research Center, Broers Building, 21 J J Thomson Avenue,
Cambridge CB3 0FA, UK}

\address{$^{2}$ Centre for Quantum Photonics, H. H. Wills Physics Laboratory
\& Department of Electrical and Electronic Engineering, University
of Bristol, Merchant Venturers Building, Woodland Road, Bristol, BS8
1UB, UK}

\address{$^{3}$University of California, Santa Barbara, Santa Barbara, California
93106, USA}

\ead{antti.niskanen@nokia.com}
\begin{abstract}
Integrated quantum photonic circuits are becoming increasingly complex.
Accurate calibration of device parameters and detailed characterisation
of the prepared quantum states are critically important for future
progress. Here we report on an effective experimental calibration
method based on Bayesian updating and Markov chain Monte Carlo integration.
We use this calibration technique to characterize a two qubit chip
and extract the reflectivities of its directional couplers. An average
quantum state tomography fidelity of 93.79$\pm$1.05$\%$ against
the four Bell states is achieved. Furthermore, comparing the measured
density matrices against a model using the non-ideal device parameters
derived from the calibration we achieve an average fidelity of 97.57$\pm$0.96$\%$.
This pinpoints non-ideality of chip parameters as a major factor in
the decrease of Bell state fidelity. We also perform quantum state
tomography for Bell states while continuously varying photon distinguishability
and find excellent agreement with theory. 
\end{abstract}

\pacs{03.67.Lx, 03.67.Hk, 42.79.-e, 42.50.Ct, 42.50.Dv, 03.65.Ta}

\maketitle
A large variety of quantum systems have been studied in recent years
for use in quantum information processing \cite{ladd,ref2,ref3,ref4}.
Among the possible implementations, quantum information processing
based on photons stands out due to its high stability, wide availability
and ease of manipulation \cite{ladd,cryto}. Recent developments in
integrated photonics have been promising from the point of view of
future large scale on-chip quantum information processing \cite{politi,SiOchip,InPchip,Sichip}.
Moreover, integrated reconfigurable quantum photonic circuits have
shown great potential for generic quantum operations \cite{jonathan,hli}.
For instance, it has been shown that the generation of arbitrary two-qubit
states and the corresponding state tomography can be realised on a
chip with high fidelity \cite{shadbolt}. However, with increasing
device complexity the accurate calibration of quantum devices becomes
a crucially important task. In this paper we study a reconfigurable
two-qubit quantum photonic device designed to create maximally entangled
states and to perform quantum state tomography on them. In particular
we focus on statistically rigorous calibration and tomography which
allow us to reach a very high fidelity between theoretically expected
and measured states.

The paper is organised as follows. We first describe the experimental
arrangement and the device design. We continue by detailing the statistical
calibration procedure and the used theoretical model that takes into
account the finite quantum interference. We also explain how the state
tomography is performed. We then proceed to the benchmarking results
for states ranging from fully mixed to nearly maximally entangled
by adjusting photon delay. We conclude by discussing the results. 

\begin{figure}[ht]
 \centering{}\includegraphics[width=0.9\textwidth]{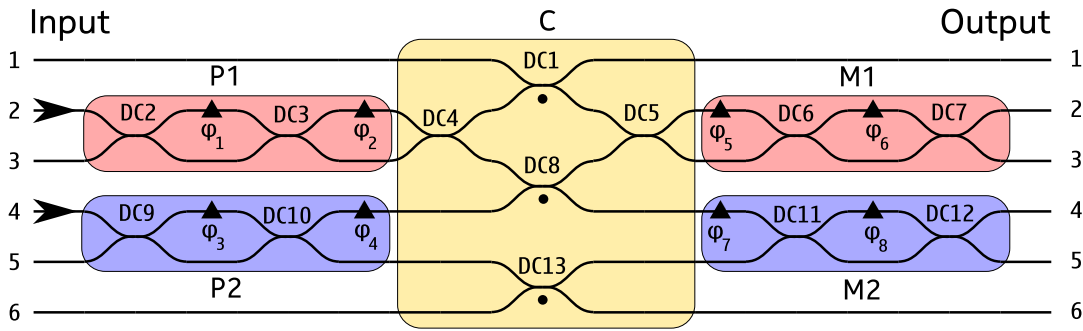} \caption{\label{fig1}Schematic design of the integrated photonic device. The
reconfigurable phase shifters are labelled by $\varphi_{1}\ldots\varphi_{8}$
and the waveguide directional couplers by DC1-13. All directional
couplers have a design reflectivity of $1/2$ except the ones marked
by a dot (reflectivity $1/3$). The shaded blocks indicate different
functions within the circuit. The pink P/M1 and blue blocks P/M2 perform
single qubit rotations on the target qubit and the control qubit,
respectively. The yellow region C performs a CNOT operation. The arrows
on the input side indicate the input waveguides for the photon pair
arriving from the source. We adopt a convention such that the control
qubit states $|0\rangle$ and $|1\rangle$ correspond to the output
channel pairs 5 and 4, respectively, while the target qubit states
$|0\rangle$ and $|1\rangle$ correspond to the output channel pairs
3 and 2, respectively.}
\end{figure}

The photonic circuit investigated here was fabricated using silica-on-silicon
technology \cite{politi}. Figure~\ref{fig1} shows a schematic design
of the device. A qubit is encoded in the amplitude and phase of a single photon
travelling on a pair of waveguides (path encoding) \cite{obrien2009,schaeff}.
For realising a two-qubit state two identical photons and four waveguides
are necessary. For instance one photon at each of the inputs 2 and
4 corresponds to the two qubit state $|10\rangle$. The chip can be
viewed as composed of three parts: the first part on the left prepares
arbitrary single qubit states (see the pink P/M1 and blue P/M2 blocks
in Fig.~\ref{fig1}). The central part (see the yellow block C in
Fig.~\ref{fig1}) is responsible for the quantum entanglement owing
to the probabilistic Controlled-NOT gate ~(CNOT) \cite{ralph}. The
blocks on the right hand side are mirror images of the preparation
blocks on the left and are used to choose the basis for projective
measurements on the two single qubits. Each block consists of a number
of directional couplers (DC) and voltage-controlled thermal variable
phase shifters~\cite{jonathan}.

The experimental setup that we used is similar to that of Ref.~\cite{hli},
the only difference being the reconfigurable photonic chip. The chip
was mounted on a chip holder and butt coupled with optical fibre V-groove
arrays at the input and output. Photon pairs were generated
by a type-I spontaneous parametric down-conversion source, pumped
using a 50 mW 405 nm laser. The 810 nm photons were filtered with
2 nm bandpass filters and collected into polarisation maintaining
fibres with aspheric lenses and then directed to the chip. The detection
of photons was done by single photon avalanche diodes connected to
the output fibre array with an efficiency of about 50$\%$. All photon
arrival times were recorded by a counting card at a time resolution
of about 165 ps. Two photon coincidences could then be detected between
every pair of output waveguides. The relative delay of the response
time between all the detectors has been calibrated and deducted before
counting coincidences. A time window of 1.5 ns was employed to count
coincidence in the experiment. This width of the window is required
because of statistic variation of detector response time. The random
two photon coincidence rate for this time window was calculated in
the experiments and was always less than 0.4\%. 

The generation, manipulation and measurement of entanglement and single
qubit mixed states has been demonstrated with a linear photonic two-qubit
chip of the same design~\cite{shadbolt} as in the present paper.
In those experiments an average quantum state tomography fidelity
of 92.8$\pm$2.5$\%$ was achieved for the four maximally entangled
Bell states. However, from experiments performed to date, it is unclear
what mechanisms cause the decrease from ideal fidelity. Possible causes
include distinguishability of the input photon pair, inaccurate phase
shifters, non ideal reflectivities of the on-chip directional couplers,
variations in the output coupling and photon detection efficiencies.
In order to make further improvement to the fidelity of chip operation
it is important to carefully characterise all these. We focus here
on the chip parameters and distinguishability.

Before using the Bayesian technique to calibrate the the chip (passive)
parameters accurate mapping of the phase shift as a function of the
applied voltage is required. The eight resistor-based variable thermo-optic
phase shifters' dependencies were measured individually with single
photons~\cite{shadbolt}. This was done using interference by sending
single photons repeatedly to the waveguides 2 or 4 and then counting
the number of output photons within a fixed time window as a function
of the heater voltage. All the eight calibrations were done while
the remaining heaters were driven at a medium power level to mimic
the conditions during a typical experiment.

With all the eight phase shifters calibrated, the reflectivities of
the 13 on-chip directional couplers $\vec{r}=(r_{1},\ldots,r_{13})$
could be determined using Bayesian inference \cite{mackay}. A statistical
model for the set of variables $X$ given the set of parameters $Y$
is described by the conditional probability $P(X|Y)$.
This model can be converted to the distribution of the
parameters $P(Y|X)$ given a set of observations $X$ using the Bayes rule 
\begin{equation}
P(Y|X)=\frac{P(X|Y)P(Y)}{P(X)},
\end{equation}
where $P(Y)$ is the prior distribution containing the initial assumptions
and $P(Y|X)$ is the posterior distribution. In the present context $X$ is the set of observed photon coincidences for different configurations whereas $Y$ is the set of device parameters.
The rule can be used iteratively for updating the distribution: The latest posterior can
be used as a prior for new data. The method makes the dependence on
underlying assumptions transparent and provides a general model that
can be applied as long as one is able to write down an appropriate
statistical model for the experiment. In the present study the required
likelihood function is for two-photon coincidences, given the parameters
of the photonic circuit. Evaluating the likelihood function exactly
is hard and we use a Markov chain Monte Carlo (MCMC) \cite{model} method
to draw samples from the posterior.

To generate the experimental data for the Bayesian inference task
we pick 1000 sets of 8 random phases uniformly generated between 0
and 2$\pi$ and bias the phase shifters accordingly. For each of the
1000 phase settings $\vec{\varphi}_{j}$, $j\in\{1,\ldots,1000\}$,
identical photons were sent to waveguides 2 and 4 while we recorded
the number of coincidences $N_{kl}^{j},\,1\leq k<l\leq6$ between
all the pairs $l$ and $k$ of the 6 output channels. That is, we
recorded the frequency of the 15 different coincidence events for
1000 randomly chosen phase settings. We denote this data set collectively
as ${\cal N}$ and the number of coincidences observed for a given
phase setting $\vec{\varphi}_{j}$ is indicated by $N^{j}=\sum_{l=1}^{6}\sum_{k=1}^{l-1}N_{kl}^{j}$.
For two channels $k\neq l$ one can express the expected probabilities
for the observed coincidences compactly as 
\begin{eqnarray}
p_{kl}= & \frac{1}{C}\left[p_{{\rm dist}}\left(|\mathcal{U}_{k2}\mathcal{U}_{l4}|^{2}+|\mathcal{U}_{l2}\mathcal{U}_{k4}|^{2}\right)\right.\label{CoincidenceProb}\\
 & \left.+(1-p_{{\rm dist}})\left|\mathcal{U}_{k2}\mathcal{U}_{l4}+\mathcal{U}_{l2}\mathcal{U}_{k4}\right|^{2}\right].\nonumber 
\end{eqnarray}
Here $p_{{\rm dist}}$ is a parameter describing the probability that
the two photons are distinguishable. $\mathcal{U}$ is the underlying
$6\times6$ unitary describing the single-photon behaviour of the
chip. $\mathcal{U}_{ik}$ can be obtained in a straightforward way
by combining the effect of the 13 directional couplers and 8 phase
shifters (see \ref{appendixA}). The probability $p_{kl}$ thus depends
on all the chip parameters of interest. The normalisation factor $C$
is obtained by summing over all the events that our detection scheme
can detect. This is needed since the events corresponding to two photons
in the same channel are not measured. For a derivation of the coincidence
probabilities starting from the device unitaries see \ref{appendixB}.
The model for the coincidence probability can be interpreted as a
statistical mixture of ideal quantum interference behaviour and distinguishable
behaviour. 

To see how this model can be used to obtain the unknown parameters,
let us consider the probability of observing the set of coincidences
${\cal N}$ given the parameters $\vec{\beta}=\left(r_{1},\ldots,r_{13},p_{{\rm dist}}\right)$.
For each of the experiments $j$ the probability of observing a number
of coincidences $N_{kl}^{j},\,(1\leq k<l\leq6)$ is given by the multinomial
distribution $N^{j}!/\left(\prod_{l=1}^{6}\prod_{k=1}^{l-1}N_{kl}^{j}!\right)\times\prod_{l=1}^{6}\prod_{k=1}^{l-1}\left(p_{kl}^{j}\right)^{N_{kl}^{j}}.$
We can therefore write the total probability as the product of 1000
multinomial distributions (for a given total number of events $N^{j}$
for each experiment $j$) 
\begin{equation}
P({\cal N}|\vec{\beta})=\prod_{j=1}^{1000}\frac{{\normalcolor N^{j}}{\Large!}}{\prod_{l=1}^{6}\prod_{k=1}^{l-1}N_{kl}^{j}!}\prod_{l=1}^{6}\prod_{k=1}^{l-1}{p_{kl}^{j}(\vec{\beta})}^{N_{kl}^{j}}.
\end{equation}
This function should be considered to be the conditional distribution
given that $N^{j}$ events have occurred for each $j$. Although $N^{j}$
is in principle a stochastic variable, it could just as well be selected
by collecting precisely that amount of data. When the $N_{kl}^{j}$
are fixed, this function of $\beta$ is called the likelihood function.
However, we were interested in the distribution of the parameters
given the observations, and not vice versa. We therefore used the
Bayes theorem to write the so-called posterior distribution for $\vec{\beta}$
as 
\begin{equation}
P(\vec{\beta}|{\cal N})\sim P({\cal N}|\vec{\beta})P_{{\rm prior}}(\vec{\beta}).
\end{equation}
Here the normalisation factor can be in principle obtained by integrating
over $\vec{\beta}$ but in practice it is hard to do without solving
for the distribution. Here we took the prior as a constant but set
it to zero in unphysical regions of the parameter space.

It is very difficult to directly evaluate $P(\vec{\beta}|{\cal N})$.
Instead, we used the Markov chain Monte Carlo Metropolis-Hastings
algorithm to perform a random walk in the parameter space $\vec{\beta}$.
This method works by starting from a random point in parameter space
and picking trial points randomly in a symmetric way. A trial point
can be accepted or rejected depending on how its probability compares
to the probability of the previous point. The new point is always
accepted if it is more probable. If the new point is less probable
it is accepted with the probability corresponding to the ratio of
probabilities. Here pseudo random numbers are used. Note that using
ratios avoids calculating the normalisation factor. Numerically it
is actually much more accurate and stable to perform the comparisons
with logarithms of the pseudo random numbers and logarithms of the
probabilities $\log(P(\vec{\beta}|{\cal N}))$. In our case this approach
avoided the need of multiplying 15000 below unity numbers before comparison.
It also eliminated the factorial term which is a constant as a function
of $\vec{\beta}$. The resulting data was stored and owing to so-called
detailed balance and ergodicity, the walk sampled the distribution
as if the points were drawn from it. One could then calculate e.g.
the moments or histograms from the data. For the data shown below
we let the system initialise for 30000 steps and we then sampled for
200000 steps. The number of steps was chosen empirically. Trials were
performed by adding a normally distributed number (standard deviation
was chosen to be 0.5\% of design/expected value) to a randomly picked
parameter. Without optimising the code the sampling takes about the
same time as the experiment using a laptop running \noun{Matlab} (overnight).

The resulting expectation values and standard deviations of the reflectivities
of the directional couplers 1-13 are shown in Table~\ref{tab1}.
The differences between fitted and designed values were expected to
be less than 5$\%$ for the process employed. The slightly larger
observed variation indicates that the fabrication process was not
fully within specifications. In addition, we found the probability
of two-photon distinguishability (which equals one minus visibility
of two photon interference) to be $p_{{\rm dist}}=4.51\pm0.11\%$.
This probability of distinguishability takes into account all contributions
that might deteriorate the visibility of two photon interference such
as non-identical spectra and polarisation of the photon source as
well as non-uniform refractive index or birefringence in the waveguides.


\begin{table}[bp]
\begin{centering}
\begin{tabular}{c>{\centering}m{2.3cm}<{\centering}>{\centering}m{4.3cm}<{\centering}r@{\extracolsep{0pt}.}l}
\hline 
 & Design values  & Fitted values  & \multicolumn{2}{m{3.0cm}}{Deviation from designed values}\tabularnewline
\hline 
$r_{1}$  & 0.3333  & 0.3257$\pm$0.0008  & -2  & 3\%\tabularnewline
$r_{2}$  & 0.5000  & 0.5186$\pm$0.0007  & 3  & 7\%\tabularnewline
$r_{3}$  & 0.5000  & 0.5063$\pm$0.0011  & 1  & 3\%\tabularnewline
$r_{4}$  & 0.5000  & 0.4494$\pm$0.0008  & -10  & 1\%\tabularnewline
$r_{5}$  & 0.5000  & 0.4526$\pm$0.0010  & -9  & 5\%\tabularnewline
$r_{6}$  & 0.5000  & 0.5375$\pm$0.0014  & 7  & 5\%\tabularnewline
$r_{7}$  & 0.5000  & 0.5635$\pm$0.0010  & 12  & 7\%\tabularnewline
$r_{8}$  & 0.3333  & 0.3175$\pm$0.0005  & -4  & 7\%\tabularnewline
$r_{9}$  & 0.5000  & 0.5381$\pm$0.0013  & 7  & 6\%\tabularnewline
$r_{10}$  & 0.5000  & 0.5009$\pm$0.0016  & 0  & 2\%\tabularnewline
$r_{11}$  & 0.5000  & 0.5204$\pm$0.0016  & 4  & 1\%\tabularnewline
$r_{12}$  & 0.5000  & 0.5760$\pm$0.0013  & 15  & 2\%\tabularnewline
$r_{13}$  & 0.3333  & 0.2967$\pm$0.0007  & -11  & 0\%\tabularnewline
\hline 
\end{tabular}
\par\end{centering}

\caption{\label{tab1}Reflectivities of the on-chip directional couplers. }
\end{table}

\begin{figure}[ht]
 \centering{}\includegraphics[width=0.9\textwidth]{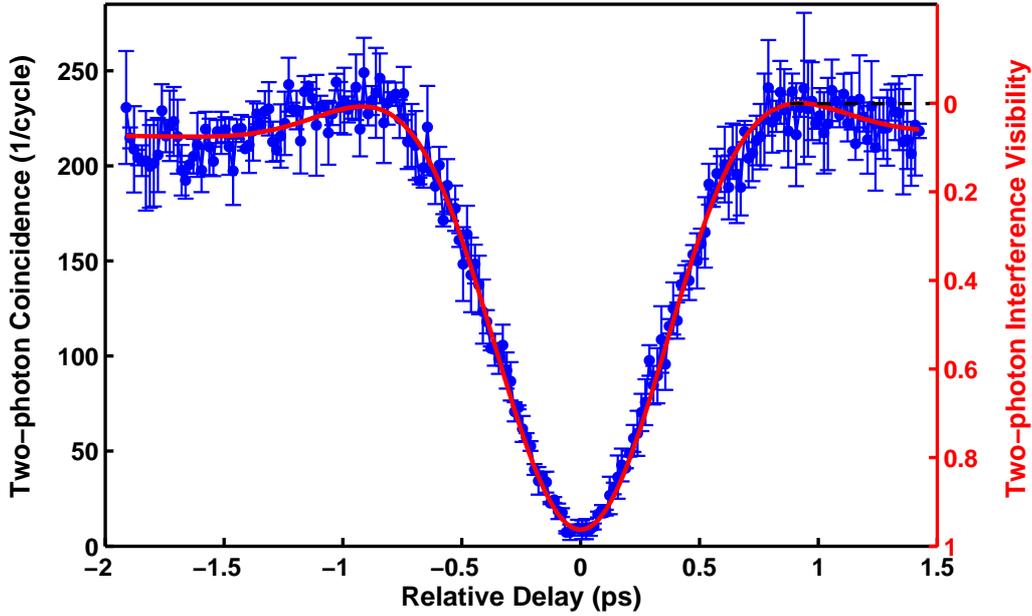} \caption{\label{fig2}Hong Ou Mandel dip scan of an on-chip Mach Zehnder interferometer
whose reflectivity is set to 0.5. The probability of distinguishability
$p_{\textnormal{dist}}$ can be obtained from a fit to the measurement
of the two photon coincidence counts. The maximum number of coincidence
counts is identified with $p_{\textnormal{dist}}=1$. }
\end{figure}

To confirm the reliability of the fitted results of both the coupler
reflectivities and distinguishability of the photons, two photon interference
experiments were carried out over various Mach Zehnder interferometers
on the chip. Figure~\ref{fig2} shows a Hong-Ou-Mandel dip~\cite{hom},
measured over the branch on the top right corner of the chip. The
Hong Ou Mandel dip was obtained by inserting photons in waveguides
1 and 4 and counting the coincidences at waveguides 2 and 3. The phase
shifters 3 and 4 on the preparation side were adjusted so that the
first interference takes place at directional coupler 5, while phase
shifters 5 and 6 were tuned so that the effective reflectivity of
the whole branch was as close to 50\% as possible (estimated 52\%).
The visibility of the dip was measured to be $96.09\pm1.8\%$. Another
Hong Ou Mandel dip measurement over the central directional coupler
8 yielded a visibility of 73.09$\pm$1.0$\%$. This is about 3.1$\%$
lower than the value (76.21$\%$) to be expected from reflectivity
of 0.3175 for coupler 8. Both of the Hong Ou Mandel scans thus resulted
in about 3-4$\%$ imperfection which is in agreement with our Markov
chain Monte Carlo-based characterisation.

Having carefully calibrated the integrated quantum photonic chip parameters
we then performed a demanding benchmark experiment. The generation
and characterization of maximally entangled Bell states offered an
ideal test case for this purpose. Using inputs 2 and 4 we prepared
the initial state $|10\rangle$ and drove the input side phase shifters
1-4 to prepare pure product states. The corresponding Bell states
$|\Phi^{-}\rangle=\frac{1}{\sqrt{2}}\left(|00\rangle-|11\rangle\right)$,
$|\Phi^{+}\rangle=\frac{1}{\sqrt{2}}\left(|00\rangle+|11\rangle\right)$,
$|\Psi^{-}\rangle=\frac{1}{\sqrt{2}}\left(|01\rangle-|10\rangle\right)$
and $|\Psi^{+}\rangle=\frac{1}{\sqrt{2}}\left(|01\rangle+|10\rangle\right)$
could then be produced utilising the CNOT gate in the centre of the
chip. However, it is clear from the characterisation of the device that
the beam splitter reflectivities deviate from their ideal values and
even for perfectly indistinguishable photons we cannot expect to be
able to prepare precisely these states. We therefore also calculated
the theoretically expected modified density matrices for the purpose
of the current benchmark test. Similarly to the Markov chain Monte
Carlo-based calibration, we modelled the theoretically expected density
matrix taking the real chip parameters into account as (see \ref{appendixC})
\begin{equation}
\rho_{\textnormal{real}}=\frac{1}{C}\left[p_{{\rm dist}}\rho_{{\rm dist}}+(1-p_{{\rm dist}})\rho_{{\rm ind}}\right]\label{TwoQubitRho}
\end{equation}
within the two-qubit subspace, where $C$ is chosen such that $\textnormal{Tr}\rho_{\textnormal{real}}=1$.
The indistinguishable part exhibiting ideal quantum interference is
obtained simply as $\rho_{{\rm ind}}=|\psi_{{\rm ind}}\rangle\langle\psi_{{\rm ind}}|$
using $|\psi_{{\rm ind}}\rangle=|\psi_{{\rm dist1}}\rangle+|\psi_{{\rm dist2}}\rangle,$
where the amplitudes of the two distinguishable possibilities are
\begin{equation}
|\psi_{{\rm dist1}}\rangle=\left(\begin{array}{c}
\mathcal{U}_{52}^{(\textnormal{p})}\mathcal{U}_{34}^{(\textnormal{p})}\\
\mathcal{U}_{52}^{(\textnormal{p})}\mathcal{U}_{24}^{(\textnormal{p})}\\
\mathcal{U}_{42}^{(\textnormal{p})}\mathcal{U}_{34}^{(\textnormal{p})}\\
\mathcal{U}_{42}^{(\textnormal{p})}\mathcal{U}_{24}^{(\textnormal{p})}
\end{array}\right)
\end{equation}
 and 
\begin{equation}
|\psi_{{\rm dist2}}\rangle=\left(\begin{array}{c}
\mathcal{U}_{32}^{(\textnormal{p})}\mathcal{U}_{54}^{(\textnormal{p})}\\
\mathcal{U}_{22}^{(\textnormal{p})}\mathcal{U}_{54}^{(\textnormal{p})}\\
\mathcal{U}_{32}^{(\textnormal{p})}\mathcal{U}_{44}^{(\textnormal{p})}\\
\mathcal{U}_{22}^{(\textnormal{p})}\mathcal{U}_{44}^{(\textnormal{p})}
\end{array}\right),
\end{equation}
where the unitary $\mathcal{U}^{(\textnormal{p})}$ is now only of
the preparation stage (see \ref{appendixA}). The distinguishable
part can be obtained as the statistical mixture of the two distinguishable
possibilities 
\begin{equation}
\rho_{{\rm dist}}=|\psi_{{\rm dist1}}\rangle\langle\psi_{{\rm dist1}}|+|\psi_{{\rm dist2}}\rangle\langle\psi_{{\rm dist2}}|.
\end{equation}
 In order to arrive at a valid density matrix the two qubit state
has to be normalised to account for the fact that the CNOT works probabilistically,
i.e. the state is projected to the two-qubit subspace. Note that the
diagonals of $\rho{}_{{\rm dist}}$ and $\rho_{{\rm ind}}$ consist
of the familiar looking elements: in the former case probabilities
are added, and in the latter case amplitudes are added.

\begin{figure}[ht]
 \centering{}\includegraphics[width=1\textwidth]{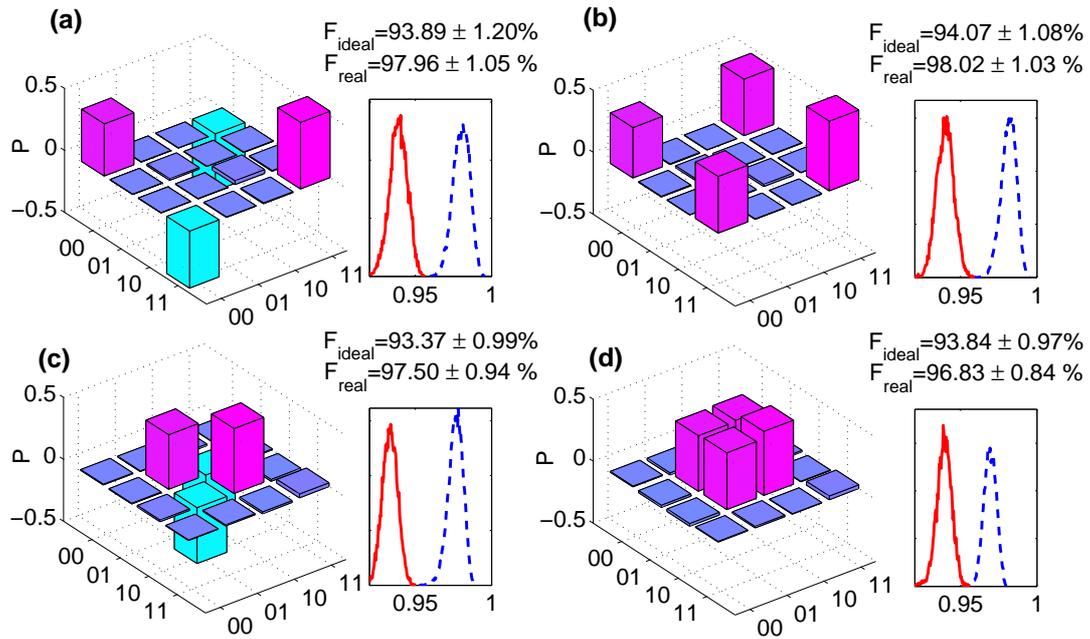} \caption{\label{fig3}Generation and measurements of Bell states. (a)-(d):
Real part of density matrix of Bell States $|\Phi^{-}\rangle$,$|\Phi^{+}\rangle$,
$|\Psi^{-}\rangle$ and $|\Psi^{+}\rangle$ respectively. The corresponding
phase shifts were $\varphi_{1}=\pi$, $\varphi_{2}=0$, $\varphi_{3}=3\pi/2$
and $\varphi_{4}=\pi/2$ for $|\Phi^{-}\rangle$; $\varphi_{1}=\pi$,
$\varphi_{2}=0$, $\varphi_{3}=\pi/2$ and $\varphi_{4}=\pi/2$ for
$|\Phi^{+}\rangle$; $\varphi_{1}=0$, $\varphi_{2}=0$, $\varphi_{3}=\pi/2$
and $\varphi_{4}=\pi/2$ for $|\Psi^{-}\rangle$; $\varphi_{1}=\pi$,
$\varphi_{2}=0$, $\varphi_{3}=3\pi/2$ and $\varphi_{4}=\pi/2$ for
$|\Psi^{+}\rangle$. The two fidelity values at the top of each figure
were calculated against the ideal Bell states and best theoretical
prediction. The quoted fidelity values indicate the Bayesian
95\% confidence intervals. The insets illustrate typical MCMC fidelity histograms for the ideal (solid) and the best theoretical expectation (dashed).}
\end{figure}

To benchmark the photonic chip we reconstructed the density matrix by
quantum state tomography \cite{QST,QST1,Blume}. Instead of the more commonly used maximum likelihood tomography we used Bayesian MCMC method. This serves two purposes: It allows us to conveniently ensure that the density matrix is physical and to obtain rigorous error bars for the density matrix fidelity against theoretical expectations. To obtain the required experimental data, we used nine different phase settings on the output side phase shifters per input state, which are in principle enough
to characterise the state fully~\cite{QST2}. That is, we measured
the qubits along $\{X,Y,Z\}\times\{X,Y,Z\}$ as accurately as possible.
Each one of the measurements gives information not only about the
corresponding two-qubit density matrix element but also about the
single-qubit terms.  However, to account for imperfections and finite
number of repetitions we resorted to numerical methods. We parametrised
the density matrix as $\rho=\sum_{r,s=0}^{4}\alpha_{rs}\sigma_{r}\sigma_{s}$,
where $\alpha_{rs}$'s are the 15 free real unknown parameters, with
$\alpha_{00}=1/4$, and $\sigma_{r}$ are the Pauli matrices including
the identity $\sigma_{0}$ in the notation. Similar to parameter estimation, we can obtain the distribution of the density matrix parameters $\vec{\alpha}$ using
$P(\vec{\alpha}|{\cal M})\sim P({\cal M}|\vec{\alpha})P(\vec{\alpha})$ or $\log P(\vec{\alpha}|{\cal M})=\log P({\cal M}|\vec{\alpha})+\log P(\vec{\alpha})$ up to a constant. Here ${\cal M}$ denotes the collective
set of observations. We can write the multinomial likelihood as
\begin{equation}
\log P({\cal M}|\vec{\alpha})=D+\sum_{i=1}^{9}\sum_{a=0}^{1}\sum_{b=0}^{1}M_{ab}^{i}\log{p_{ab}^{i}},
\end{equation}
where D is a constant, the index $i$ runs over the nine tomography
phase settings, $M_{ab}^{i}$ is the number of times that we detected
the qubit state $|ab\rangle$ and $p_{ab}^{i}$ is the corresponding
expected probability. These probabilities depend on both the unknown
density matrix parameters that we optimise over and the currently known
phase settings of the shifters 5-8. We used a uniform prior $P(\vec{\alpha})$ over physical density matrices, i.e. $P(\vec{\alpha})$ is constant whenever the corresponding eigenvalues of $\rho$ are non-negative and zero otherwise. In practice this means that $\log P(\vec{\alpha})$ is set to a large value (ideally infinite) whenever the MCMC algorithm attempts to move outside the physical region. There are numerous other alternative ways to choose the prior \cite{prior1,prior2}. We chose the present one due to numerical convenience.
The real part of measured density matrices of all
the four Bell states are shown in Figure~\ref{fig3}. The fidelity estimates and the mean density matrices are obtained by averaging over the random walk. Both the fidelity
against the ideal Bell state and the fidelity against the predicted
density matrix taking into account the calibrated parameters are shown.
These are denoted by $F_{{\rm ideal}}$ and $F_{{\rm real}}$, respectively.
We used the definition of fidelity \cite{jozsa}
\begin{equation}
F_{\textnormal{ideal/real}}=\left(Tr\sqrt{\sqrt{\rho_{\textnormal{ideal/real}}}\rho_{\textnormal{exp}}\sqrt{\rho_{\textnormal{ideal/real}}}}\right)^{2},
\end{equation}
where $\rho_{{\rm exp}}$ is the experimentally measured density matrix.
To calculate the fidelity $F_{\textnormal{ideal}}$ we choose the
reference density matrix $\rho_{\textnormal{ideal}}$ to be the density
matrix of one of the Bell states. To obtain $F_{\textnormal{real}}$
we set the reference density matrix to the density matrix calculated
using Eq.~(\ref{TwoQubitRho}) with the same settings for the phase
shifters as for the Bell states but using real coupler efficiencies
and taking the non-ideal photon indistinguishability into account.
For the four Bell states $F_{{\rm ideal}}$ is 93.79$\pm$1.05$\%$
on average. This is an improvement over previously reported \cite{shadbolt}
fidelities. The remaining 5-6$\%$ imperfection mostly arises from
the non ideal reflectivities and the photon source, as the average
fidelity increases to 97.57$\pm$0.96$\%$ for $F_{{\rm real}}$.
This indicates that a major source of the decrease of the fidelity
are the non-ideal reflectivities of the on-chip directional couplers.

\begin{figure}[ht]
 \centering{}\includegraphics[width=1\textwidth]{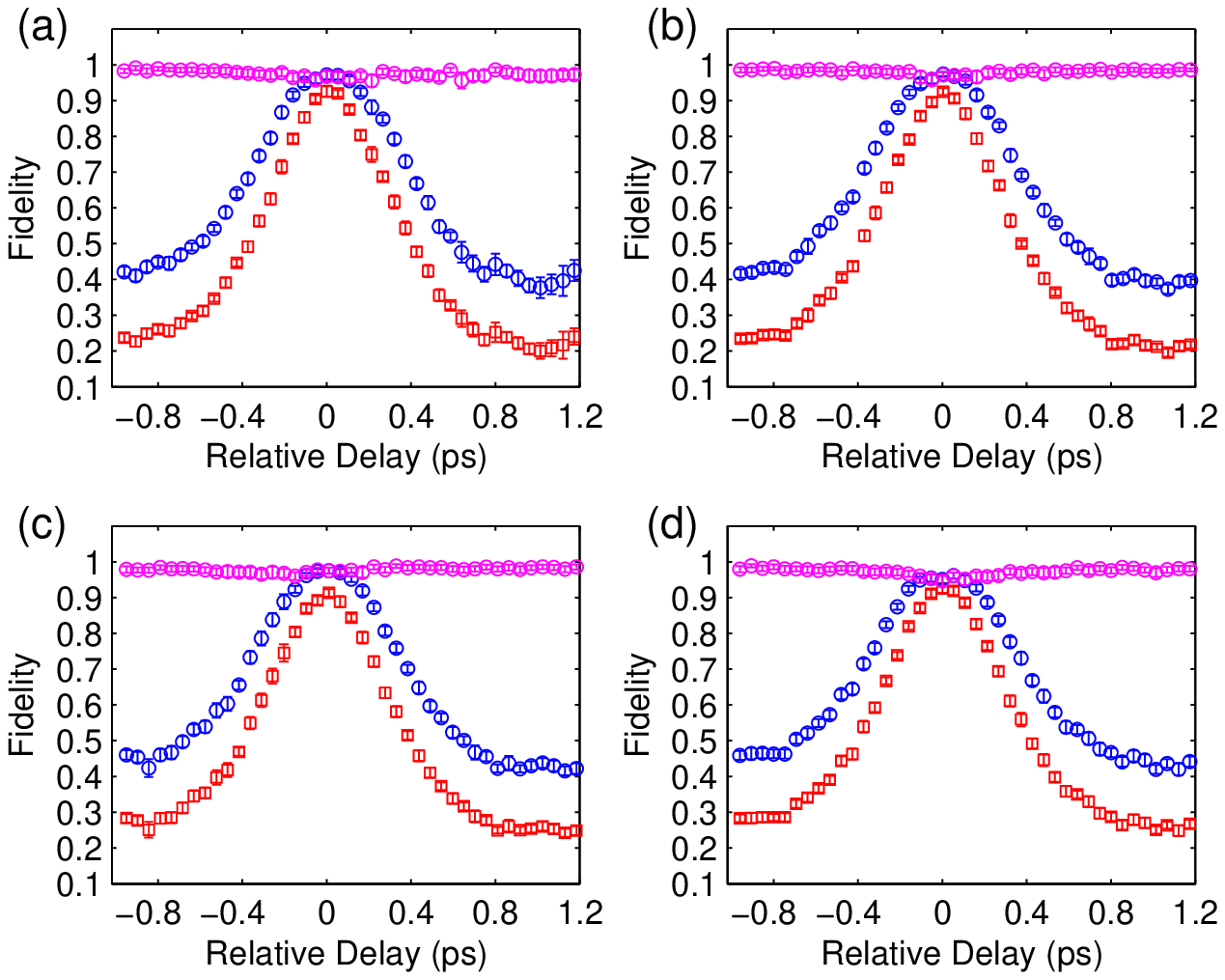} \caption{\label{fig4}Bell states' fidelity as a function of the relative delay
of the two photons. Red: $F_{{\rm ideal}}$, fidelity against ideal
Bell states. Blue: $F_{{\rm real}}$, fidelity against modified expectation
with fitted reflectivities and probability of distinguishability.
Magenta: fidelity against the best theoretical expectation using fitted
reflectivities and the delay-dependent probability of distinguishability.
This delay-dependent probability is obtained from the independent
Hong Ou Mandel dip scan in Fig.~\ref{fig2}.}
\end{figure}

To further demonstrate the agreement with the theoretical model and
the experimental results, we varied the relative delay of the two
photons and performed quantum state tomography for each fixed delay.
This experiment can be viewed as a Hong-Ou-Mandel measurement for
Bell states. Depending on the delay between the two photons, the four
two-qubit states were expected to change between maximally entangled
and totally mixed as predicted by the model presented above. We compared
the measured density matrices against (i) the ideal Bell states, (ii)
against the best expected density matrix with finite but optimal $p_{{\rm dist}}$
with real reflectivities and (iii) finally against the delay dependent
$\rho(p_{{\rm dist}})$ (with real reflectivities), where $p_{{\rm dist}}$
was deduced from the independent measurement in Fig.~\ref{fig2}.
Figure~\ref{fig4} (a)-(d) illustrate the results for $|\Phi^{-}\rangle$,
$|\Phi^{+}\rangle$, $|\Psi^{-}\rangle$ and $|\Psi^{+}\rangle$,
respectively. There are three curves for each state: the lowest (red)
curve corresponds to case (i), the middle (blue) corresponds to case
(ii) and top (magenta) corresponds to case (iii). In cases (i) and
(ii) one can clearly see how the fidelities peak in analogy with
the Hong-Ou-Mandel effect. Comparing the experiments with the detailed
model results in an increase in the fidelities. The tops of the four 
peaks correspond to Fig.~\ref{fig3}. Case (iii) shows the agreement
of theory and experiment most clearly; the fidelity of the measured
density matrices as a function of delay agrees almost perfectly with
the model for the delay dependent $\rho_{\textnormal{real}}(p_{\textnormal{dist}})$
which is a statistical mixture of distinguishable and indistinguishable
behaviour. How the reconstructed density matrices evolve as a function
of the delay is shown in a supplementary video.

In conclusion, we have presented an accurate calibration method for
a reconfigurable two-qubit quantum photonic chip based on Bayesian
updating and Markov chain Monte Carlo simulation. An alternative approach to tomography of linear optics circuits has been presented in \cite{Superstable}. Our method differs in several
points. Additional to employing a statistical inference method we
use an explicit model of the device and do not look for a general
unitary. This reduces the number of parameters that need to be estimated.
Furthermore, our method introduces minimal disturbance from calibration
to the actual measurements as it is not necessary to switch the input
ports for the photons. 
Having obtained the real chip and
source parameters we checked the fidelity of a prepared state against
a theoretical prediction using these parameters. A maximum fidelity
of 98.02$\pm$1.03$\%$ against the theory prediction using the real
device parameters shows that practically all error sources are accounted
for and improving the photon source as well as the chip fabrication
will make it possible to prepare states with fidelities larger than
$99\%$. A remarkable fact is that the almost unit fidelity also extends
to mixed states. This enables us to reliably prepare not only maximally
entangled states but also mixed states with a varying degree of mixedness.
A source of such states can be of considerable interest in investigating
mixed state quantum computation and test concepts like quantum discord.

\appendix

\section{Single photon unitary\label{appendixA}}

The wavefunction of a single photon travelling through the photonic
chip can be described in the waveguide basis 
\begin{equation}
\left|\psi\right\rangle =\sum_{i}c_{i}\left|i\right\rangle 
\end{equation}
 where $\left|i\right\rangle $ denotes the presence of a photon in
waveguide $i=1,\ldots,6$. The photon travels through a series of
phaseshifters $S$ and directional couplers $D$. The action of each
linear element on the wavefunction can be described by a 6 x 6 matrix
such that the wavefunction after the element is 
\begin{equation}
\left|\psi\right\rangle =\sum_{ik}X_{ik}c_{k}\left|i\right\rangle ,\quad X=D,S.
\end{equation}
 A phaseshifter applying the phase shift $\phi$ to the photon in
waveguide $a$ is described by the matrix 
\begin{equation}
S_{kl}^{(a)}(\phi)=\delta_{kl}\left[\delta_{ak}e^{i\phi}+(1-\delta_{ak})\right],
\end{equation}
where $k$ and $l$ label the waveguides. For example a phaseshifter
on waveguide 2 is given by 
\begin{equation}
S^{(2)}(\phi)=\left(\begin{array}{cccccc}
1 & 0 & 0 & 0 & 0 & 0\\
0 & e^{i\phi} & 0 & 0 & 0 & 0\\
0 & 0 & 1 & 0 & 0 & 0\\
0 & 0 & 0 & 1 & 0 & 0\\
0 & 0 & 0 & 0 & 1 & 0\\
0 & 0 & 0 & 0 & 0 & 1
\end{array}\right),
\end{equation}
which is a unity matrix, where the 2nd element on the diagonal is
replaced by a phase factor. A directional coupler between waveguide
$a$ and $b$ with reflectivity $r$ is described by 
\begin{eqnarray}
D_{kl}^{(a,b)}(r) & = & \left(\delta_{ak}+\delta_{bk}\right)\left(\delta_{al}+\delta_{bl}\right)\left[\sqrt{r}\delta_{kl}+i\sqrt{1-r}\left(1-\delta_{kl}\right)\right]\nonumber \\
 &  & +\left(1-\delta_{ak}-\delta_{bk}\right)\delta_{kl}.
\end{eqnarray}
As an example a directional coupler with reflectivity $r$ between
waveguide 2 and waveguide 3 is given by 
\begin{equation}
D^{(2,3)}(r)=\left(\begin{array}{cccccc}
1 & 0 & 0 & 0 & 0 & 0\\
0 & \sqrt{r} & i\sqrt{1-r} & 0 & 0 & 0\\
0 & i\sqrt{1-r} & \sqrt{r} & 0 & 0 & 0\\
0 & 0 & 0 & 1 & 0 & 0\\
0 & 0 & 0 & 0 & 1 & 0\\
0 & 0 & 0 & 0 & 0 & 1
\end{array}\right).
\end{equation}
 The unitary of the chip $\mathcal{U}$ has a rather complicated form
and we will therefore present it as the product of its elementary
building blocks. It can be written in terms of the preparation stage
unitary $\mathcal{U}^{(p)}$ and the unitary describing the selection
of the measurement bases $\mathcal{U}^{(m)}$ as 
\begin{equation}
\mathcal{U}=\mathcal{U}^{(\textnormal{m})}\cdot\mathcal{U}^{(\textnormal{p})}
\end{equation}
 where the preparation stage unitary can be broken down further into
\begin{equation}
\mathcal{U}^{(\textnormal{p})}=\mathcal{U}_{C}\cdot\mathcal{U}_{P1}\cdot\mathcal{U}_{P2}
\end{equation}
 with the unitary of the central part given by 
\begin{equation}
\mathcal{U}_{C}=D^{(2,3)}(r_{5})\cdot D^{(5,6)}(r_{13})\cdot D^{(3,4)}(r_{8})\cdot D^{(1,2)}(r_{1})\cdot D^{(2,3)}(r_{4}),
\end{equation}
 where the reflectivities $r_{i}$ (design and measured) can be found
in Tab~\ref{tab1}. The two other blocks are given by 
\begin{equation}
\mathcal{U}_{P1}=S^{(2)}(\varphi_{2})\cdot D^{(2,3)}(r_{3})\cdot S^{(2)}(\varphi_{1})\cdot D^{(2,3)}(r_{2})
\end{equation}

\begin{equation}
\mathcal{U}_{P2}=S^{(4)}(\varphi_{4})\cdot D^{(4,5)}(r_{10})\cdot S^{(4)}(\varphi_{3})\cdot D^{(4,5)}(r_{9})
\end{equation}
 In the same way the unitary describing the selection of the measurement
bases can be written as 
\begin{equation}
\mathcal{U}^{(\textnormal{m})}=\mathcal{U}_{M1}\cdot\mathcal{U}_{M2}
\end{equation}
 with 
\begin{equation}
\mathcal{U}_{M1}=D^{(2,3)}(r_{7})\cdot S^{(2)}(\varphi_{6})\cdot D^{(2,3)}(r_{6})\cdot S^{(2)}(\varphi_{5})\cdot
\end{equation}

\begin{equation}
\mathcal{U}_{M2}=D^{(4,5)}(r_{12})\cdot S^{(4)}(\varphi_{8})\cdot D^{(4,5)}(r_{11})\cdot S^{(4)}(\varphi_{7})\cdot
\end{equation}
 This fully defines the single photon unitary of the photonic chip
under consideration.

\section{\label{appendixB}From Unitaries to Probabilities}

Once the unitary is known we can proceed to calculate the probabilities
for coincidence counts at the detectors. We will assume that the source
produces photon with a given frequency distribution, so that a photon
in wave guide $i$ is described by 
\begin{equation}
\left|i\right\rangle =\int d\omega\,\alpha^{*}(\omega)\hat{a}_{i}^{\dagger}(\omega)\left|0\right\rangle ,
\end{equation}
where $\left|0\right\rangle $ is the vacuum state, $\hat{a}_{i}^{\dagger}(\omega)$
a creation operator for a photon with frequency $\omega$ in waveguide
$i$ and $\alpha(\omega)$ a normalised amplitude for a frequency
so that $\int d\omega\mbox{\ensuremath{\left|\alpha(\omega)\right|}}^{2}=1$.
With a photonic circuit described by the single photon unitary $\mathcal{U}$
and an initial state with two photons entering the waveguides $m$
and $n$, $n\neq m$, we can write the wave function at the exit of
the circuit as
\begin{equation}
\left|\Psi\right\rangle =\sum_{ij}\mathcal{U}_{im}\mathcal{U}_{jn}\int d\omega_{1}d\omega_{2}\,\alpha_{1}^{*}(\omega_{2})\alpha_{2}^{*}(\omega_{1})\left|i\omega_{1},\, j\omega_{2}\right\rangle ,\label{LargePsi}
\end{equation}
where we defined the two photon state
\begin{equation}
\left|i\omega_{1},\, j\omega_{2}\right\rangle =\hat{a}_{j}^{\dagger}(\omega_{2})\hat{a}_{i}^{\dagger}(\omega_{1})\left|0\right\rangle \label{TwoPhotonState}
\end{equation}
The coincidence counts at detectors $k$ and $l$, $k\neq l$, are
then given by
\begin{equation}
n_{kl}=\left\langle \Psi\right|\hat{N}_{k}\hat{N}_{l}\left|\Psi\right\rangle ,
\end{equation}
where the number operators are given by
\begin{equation}
\hat{N}_{k}=\int d\omega\,\hat{a}_{k}^{\dagger}(\omega)\hat{a}_{k}(\omega).
\end{equation}
After a lengthy but straightforward calculation, contracting pairs
of creation and annihilation operators, using the commutation relation
\begin{equation}
\left[\hat{a}_{i}(\omega_{2}),\hat{a}_{j}^{\dagger}(\omega_{1})\right]=\delta_{ij}\delta(\omega_{1}-\omega_{2})
\end{equation}
we obtain
\begin{eqnarray}
n_{kl} & = & \left(\left|\mathcal{U}_{kn}\right|^{2}\left|\mathcal{U}_{lm}\right|^{2}+\left|\mathcal{U}_{km}\right|^{2}\left|\mathcal{U}_{ln}\right|^{2}\right)p_{\textnormal{dist}}\nonumber \\
 & + & \left|\mathcal{U}_{kn}\mathcal{U}_{lm}+\mathcal{U}_{ln}\mathcal{U}_{km}\right|^{2}\left(1-p_{\textnormal{dist}}\right)
\end{eqnarray}
with 
\begin{equation}
p_{\textnormal{dist}}=\int d\omega_{1}d\omega_{2}\,\alpha_{1}(\omega_{1})\alpha_{1}^{*}(\omega_{2})\alpha_{2}(\omega_{2})\alpha_{2}^{*}(\omega_{1}).\label{pdist}
\end{equation}
Note that we have here selected the frequency distribution to distinguish
the photons but a similar argument can be made for other degrees of
freedom, e.g. polarisation. In our experiments we are interested in
the the probability of a certain coincidence, given that we insert
a photon into each of the waveguides 2 and 4. We obtain the probability
of coincidence as $p_{kl}=n_{kl}/C$
\begin{eqnarray}
p_{kl} & = & \frac{1}{C}\left(\left|\mathcal{U}_{k2}\right|^{2}\left|\mathcal{U}_{l4}\right|^{2}+\left|\mathcal{U}_{k4}\right|^{2}\left|\mathcal{U}_{l2}\right|^{2}\right)p_{\textnormal{dist}}\nonumber \\
 & + & \frac{1}{C}\left|\mathcal{U}_{k2}\mathcal{U}_{l4}+\mathcal{U}_{k4}\mathcal{U}_{l2}\right|^{2}\left(1-p_{\textnormal{dist}}\right)
\end{eqnarray}
with the normalisation constant given by 
\begin{equation}
C=\sum_{k=1}^{6}\sum_{l=1}^{k-1}n_{kl},
\end{equation}
which is exactly the form of Eq.~(\ref{CoincidenceProb}). 

We now show that $p_{\textnormal{dist}}$ can be directly obtained
from a Hong-Ou-Mandel type experiment. In such a measurement two photons
are directed to two different ports of a directional coupler with
reflectivity $r$=1/2. The unitary is given by 
\begin{equation}
\mathcal{U}=\left(\begin{array}{cc}
\sqrt{r} & i\sqrt{1-r}\\
i\sqrt{1-r} & \sqrt{r}
\end{array}\right).
\end{equation}
What is measured is the number of coincidence counts relative to the
number of counts obtained for completely distinguishable photons.
For a beam splitter with reflectivity $r$ we obtain 
\[
n_{12}=(1-p_{\textnormal{dist}})(1-2r)^{2}+p_{\textnormal{dist}}(1-2r+2r^{2}).
\]
Normalising this to the number of coincidence counts for completely
distinguishable photons we get 
\[
x=\frac{(1-p_{\textnormal{dist}})(1-2r)^{2}+p_{\textnormal{dist}}(1-2r+2r^{2})}{1-2r+2r^{2}}
\]
or expanding around $r=1/2+\delta r$ to second order in $\delta r$
we obtain
\[
x=p_{\textnormal{dist}}+8(1-p_{\textnormal{dist}})\delta r^{2}.
\]
This shows that we can directly read off the function $p_{\textnormal{dist}}$
from a Hong-Ou-Mandel type experiment and that the error induced by
using a non-ideal beam splitter is quadratic in the deviation.

\section{\label{appendixC}The Two Qubit Density Matrix}

If we want to consider the photonic quantum circuit in the light of
quantum computation we have to assign the meaning of qubits to certain
combinations of photons. We do this in the following way: Waveguides
2 and 3 encode one qubit in dual rail encoding, while waveguides 4
and 5 encode the second one. A photon present in waveguide 2 and the
second photon in waveguide 4 maps onto the logical two qubit state
$\left|11\right\rangle _{L}$. Similarly we make the mappings $\left|25\right\rangle \rightarrow\left|01\right\rangle _{L}$,
$\left|34\right\rangle \rightarrow\left|10\right\rangle _{L}$ and
$\left|35\right\rangle \rightarrow\left|00\right\rangle _{L}$. Given
the full wave function on the chip and the fact that the detectors
only register the presence of a photon, but do not distinguish parameters
like polarisation or wavelength, we can construct a reduced density
matrix for the two qubit subspace as
\begin{equation}
\rho=\frac{1}{C}\sum_{(mn),(kl)}\rho_{mn,kl}\left|mn\right\rangle \left\langle kl\right|,\label{dmatr}
\end{equation}
where the pairs $kl$ and $mn$ are taken from $\{35,25,34,24\}.$
The coefficients are given by
\begin{equation}
\rho_{mn,kl}=\int d\omega_{1}d\omega_{2}\left\langle m\omega_{1},\, n\omega_{2}\right|\left.\Psi\right\rangle \left\langle \Psi\right|\left.k\omega_{1},\, l\omega_{2}\right\rangle ,
\end{equation}
with $\left|\Psi\right\rangle $ given by Eq.~(\ref{LargePsi}) and
$\left|k\omega_{1},\, l\omega_{2}\right\rangle $ defined in Eq.~(\ref{TwoPhotonState})
and the partial trace running over the frequency degree of freedom.
In a similar calculation to the previous section and assuming that
single photons are inserted in waveguide 2 and 4 we obtain
\begin{eqnarray}
\rho_{mn,kl} & = & \left(\mathcal{U}_{m2}\mathcal{U}_{n4}\mathcal{U}_{k2}^{*}\mathcal{U}_{l4}^{*}+\mathcal{U}_{n2}\mathcal{U}_{m4}\mathcal{U}_{l2}^{*}\mathcal{U}_{k4}^{*}\right)p_{\textnormal{dist}}\nonumber \\
 &  & +\left(\mathcal{U}_{m2}\mathcal{U}_{n4}+\mathcal{U}_{n2}\mathcal{U}_{m4}\right)\left(\mathcal{U}_{k2}^{*}\mathcal{U}_{l4}^{*}+\mathcal{U}_{l2}^{*}\mathcal{U}_{k4}^{*}\right)\left(1-p_{\textnormal{dist}}\right)
\end{eqnarray}
with $p_{dist}$ defined in Eq.~(\ref{pdist}). Reinserting this
in the equation for the density matrix Eq.~(\ref{dmatr}) we see
that we can define
\begin{equation}
\left|\psi_{\textnormal{dist1}}\right\rangle =\sum_{(mn)}\mathcal{U}_{n2}\mathcal{U}_{m4}\left|mn\right\rangle 
\end{equation}
and

\begin{equation}
\left|\psi_{\textnormal{dist2}}\right\rangle =\sum_{(mn)}\mathcal{U}_{m2}\mathcal{U}_{n4}\left|mn\right\rangle 
\end{equation}
so that 
\begin{eqnarray}
\rho & = & \frac{p_{\textnormal{dist}}}{C}\left(\left|\psi_{\textnormal{dist1}}\right\rangle \left\langle \psi_{\textnormal{dist1}}\right|+\left|\psi_{\textnormal{dist2}}\right\rangle \left\langle \psi_{\textnormal{dist2}}\right|\right)\nonumber \\
 & + & \frac{1-p_{dist}}{C}\left(\left|\psi_{\textnormal{dist1}}\right\rangle +\left|\psi_{\textnormal{dist2}}\right\rangle \right)\left(\left\langle \psi_{\textnormal{dist1}}\right|+\left\langle \psi_{\textnormal{dist2}}\right|\right).
\end{eqnarray}
This is the form of the two qubit density matrix given in Eq.~(\ref{TwoQubitRho}).

\pagebreak{} 

\begin{thebibliography}{10}
\bibitem{ladd} Ladd T D \etal 2010\emph{ Nature} \textbf{464} 45

\bibitem{ref2} Criger B, Passante G, Park D and Laflamme R 2012 \textit{Phil.
Trans. R. Soc. }\textit{\emph{A}} \textbf{370} 4620-35

\bibitem{ref3} M. Hofheinz \textit{et al. 2009} \emph{Nature} \textbf{459}
546

\bibitem{ref4} Barreiro J T \etal 2011 \textit{Nature} \textbf{470}
486

\bibitem{cryto} Gisin N, Ribordy G, Tittel W, Zbinden H 2002 \textit{Rev.
Mod. Phys.} \textbf{74} 145

\bibitem{politi} Politi A, Cryan M J, Rarity J G, Yu S and O'Brien
J L 2008 \textit{Science} \textbf{320} 646

\bibitem{SiOchip} Smith B J, Kundys D, Thomas-Peter N, Smith P G
and Walmsley I A 2009 \textit{Opt. Express} \textbf{17} 13516

\bibitem{InPchip} Coldren L A \etal 2011 \textit{J. Lightwave Technol.}
\textbf{29} 554

\bibitem{Sichip} Vlasov Y, Green W M and Xia F 2008 \textit{Nature
Photon.} \textbf{2} 242

\bibitem{jonathan} Matthews J C F, Politi A, Stefanov A and O'Brien
J L 2009 \textit{Nature Photon.} \textbf{3} 346

\bibitem{hli} Li H W \etal 2011 \textit{New J. Phys.} \textbf{13}
115009

\bibitem{shadbolt} Shadbolt P J \etal 2012 \textit{Nature Photon.}
\textbf{6} 45

\bibitem{Superstable}Laing A, O'Brien J L 2012 \emph{Preprint} arXiv:1208.2868

\bibitem{mackay} MacKay D J 2003 \textit{Information Theory, Inference
and Learning Algorithms} (Cambridge: Cambridge University Press)

\bibitem{obrien2009} O'Brien J L, Furusawa A and Vukovic J 2009 \textit{Nature
Photon.} \textbf{3} 687

\bibitem{schaeff} Schaeff C, Polster R, Lapkiewicz R, Fickler R,
Ramelow S and Zeilinger A 2012 \textit{Opt. Express} \textbf{20} 16145

\bibitem{ralph} Ralph T C, Langford N K, Bell T B and White A G 2002
\textit{Phys. Rev. }\textit{\emph{A}} \textbf{65} 062324

\bibitem{model} Hastie T, Tibshirani R and Friedman J 2009 \textit{The
Elements of Statistical Learning} (New York: Springer-Verlag)

\bibitem{hom} Hong C K, Ou Z Y and Mandel L 1987 \textit{Phys. Rev.
Lett.} \textbf{59} 2044

\bibitem{QST}Banaszek K, D\textquoteright{}Ariano G M, Paris M G
A and Sacchi M F 1999 \textit{Phys. Rev. }\textit{\emph{A}} \textbf{59}
010304(R)

\bibitem{QST1} Cramer M \etal 2010 \textit{Nat. Commun.} \textbf{1}
149

\bibitem{Blume} Blume-Kohout R 2010 \textit{New J. Phys.} \textbf{12} 043034
\bibitem{QST2} Altepeter J B, Jeffrey E R and Kwiat P G 2006 \textit{Adv.
At. Mol. Opt. Phy.} \textbf{52} 105 (Editors: Berman P and Lin C, Academic Press, Elsevier)

\bibitem{prior1} Yin J O S and van Enk S J 2011 \textit{Phys. Rev. A} \textbf{ 83} 022326
\bibitem{prior2} Osipov V A, Sommers H-J and Zyczkowski K 2010 \textit{J. Phys. A} \textbf{ 43} 055302
\bibitem{jozsa} Jozsa R 1994 \textit{J Mod Optics} \textbf{41} 2315

\end{thebibliography}
\end{document}